\documentclass[a4paper,12pt]{article}
\usepackage{setspace}
\usepackage{amsmath}
\usepackage{graphicx}
\usepackage[square,numbers,sort&compress]{natbib}
\title{Dynamics of nano tippe top}
\author{Yue Chan\footnote{Corresponding author's email address: yc321@uow.edu.au}
, Ngamta Thamwattana and James M. Hill\\
\footnotesize{Nanomechanics Group, School of Mathematics and Applied Statistics,}\\
\footnotesize{University of Wollongong, Wollongong, NSW 2522,
Australia}} 

\begin{document}

\maketitle

\abstract{We investigate the motion of  a nano tippe top, which is
formed from a C$_{60}$ fullerene, and which is assumed to be
spinning on either a graphene sheet or the interior of a
single-walled carbon nanotube. We assume no specific geometric
configuration for the top, however for example, the nano tippe top
might be formed by joining a fullerene C$_{60}$ with a small segment
of a smaller radius carbon nanotube. We assume that it is spinning
on a graphene sheet or a carbon nanotube surface only as a means of
positioning and isolating the device, and the only effect of the
graphene or the carbon nanotube surface is only through the
frictional effect generated at the point of the contact. We employ
the same basic physical ideas originating from the classical tippe
top and find that the total retarding force, which comprises both a
frictional force and a magnetic force at the contact point between
the C$_{60}$ fullerene and the graphene sheet or the inner surface
of the single-walled carbon nanotube, induces the C$_{60}$ molecule
to spin and precess from a standing up position to a lying down
position. Unlike the classical tippe top, the nanoscale tippe top
does not flip over since the gravitational effect is not sufficient
at the nano scale. After the precession, while the molecular top
spins about its lying down axis, if we apply the opposite retarding
magnetic force at the contact point, then the molecule will return
to its standing up position. The standing up and the lying down
configurations of the nano tippe top during the precession and
retraction processes demonstrate its potential use as a memory
device in nano-computing.

\maketitle
\section{Introduction}

The discovery of fullerenes \cite{Smalley} and carbon nanotubes
\cite{Iijima} has led to numerous studies on their properties and
their various potential applications in nano devices.  In this
paper, we focus on the mechanics of a nanoscale tippe top comprising
a fullerene C$_{60}$ which for example might be joined to a carbon
nanotube and is spinning on a graphene sheet or inside a large
single-walled carbon nanotube. We assume that the nano tippe top is
at the equilibrium configuration either on the graphene sheet or
inside the outer carbon nanotube and that the spinning occurs due to
the application of an external magnetic field. We refer to this
structure as a nano tippe top or simply as a nano top. We note that
in the case of the top spinning inside a nanotube, that although the
effect of the outer carbon nanotube is not directly incorporated
into our calculations, we have in mind that the outer carbon
nanotube acts as a barrier between the nano top and the external
environment.

The flip over of the classical tippe top has attracted much
attention due to its ``apparent" violation of the principle of
conservation of energy during the top's inversion (i.e. the rise in
its center of mass results in the sudden increase in its potential
energy but apparently no gain in other energies) as well as the lack
of a complete mathematical description of this inversion phenomenon
\cite{Braams, Hugenholtz, OR, Leutwyler, Ebenfeld, Gray, Bou}.
However, Cohen \cite{Richard} provides an analysis and a numerical
study on the tippe top. He views the tippe top as an eccentric
sphere, for which its center of mass is different from its geometric
center, and incorporates into his model Coulomb friction at the
contact point to describe the tippe top's motion. For example, this
eccentric sphere can be manufactured either by creating the top with
two different mass densities or by puncturing the sphere and
inserting a stem. At the nano scale, the latter method may be
achieved by introducing a defect on the surface of a fullerene
C$_{60}$ and then joining to this defect a small segment of a carbon
nanotube. We refer the reader to Nasibulin~{\em et al.}
\cite{Nasibulin} for the possible creation of such nano tippe tops.
Classically, the friction between the eccentric sphere and the
relatively rough surface plays a vital role for the top's inversion,
and the sudden gain in the potential energy during the inversion
results from the loss in the top's rotational kinetic energy, which
can also be observed from the reduction in the spinning of the
inverted top. Furthermore, since gravity and the normal force act
solely along the axis fixed in space and provide zero torque to the
top, the friction, which acts offset from the space-fixed axis and
against the rotational motion of the top, and is therefore the only
source providing an external torque to slow the spinning of the top
down \cite{Braams}. In addition, Ueda \emph{et al.} \cite{Ueda} show
theoretically that the top's initial spinning and the ratio of the
top's two different principal moments of inertia at the center of
mass play an important role to determine the top's inversion. For
example, the flip over phenomenon can only occur when its initial
spinning is above a certain threshold. Since friction at the nano
scale is ultra-low \cite{Falvo, Kolmogorov}, our numerical results
show that the friction itself is insufficient to make the nano tippe
top precessing about the fixed-body axis. Therefore, in this paper
we introduce a sufficiently large retarding magnetic force at the
contact point to act analogously to the effect of friction. However,
we find numerically that instead of flipping over, the nano top
prefers to spin in a stable lying down configuration, which suggests
that the effect of gravity is negligible at the nano scale. In
particular, the nano top behaves more like a hard-boiled egg
spinning on a rough surface \cite{Moffatt}.

As mentioned  above, to induce the spinning effect of a nano top we
introduce a retarding magnetic force at the contact point between
the fullerene C$_{60}$ and the carbon nanotube's wall. We find from
Wood \emph{et al.} \cite{Wood} who experimentally produce
ferromagnetic fullerenes C$_{60}$  that the magnetically strongest
fullerenes are formed at 800 K with the magnetic moment per molecule
of 0.38$\mu_{B}$, where $\mu_B $ denotes the Bohr magneton constant.
This result indicates that we can initiate the spinning of a
fullerene in a preferred direction by applying an external magnetic
field to the center of the fullerene and similarly at the contact
point for generating the retarding magnetic force.

In the following section, we briefly state vector equations which
are used to describe motion of a tippe top. We note that explicit
forms of these equations are given in Appendix A. In Section~III, we
provide numerical results for a nano tippe top, which is driven by a
constant magnetic force in both $x$- and $y$-directions. We verify
our numerical schemes by examining the classical tippe top, as
presented in Appendix~A.  Conclusion of the paper is given in
Section~IV. In Appendix B, we use the basic equations as given in
Appendix~A to study the stability  for certain configurations of a
nano tippe top and in particular Appendix~C considers the
compatibility between our numerical results and asymptotic
expansions when the nano top is in the lying position. Finally,
while Section~III considers $H_x=H_y=H$, in Appendix~D we assume a
magnetic force which is applied only in the $y$-direction and only
for a finite time $t_0$.

\section{Equations of motion}

\begin{figure}[!h]
\begin{flushleft}
    \includegraphics[width=8.5cm,height=5.6cm]{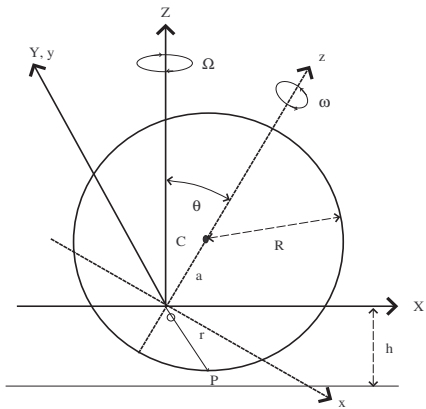}
\end{flushleft}
  \caption{\footnotesize{Schematic of nano tippe top}}
  \label{fig:Figure1}
\end{figure}
In this section, we state the equations of motion for a nano tippe
top, which is schematically illustrated in Fig.~\ref{fig:Figure1}.
Owing to the axially symmetry of the density of the nano top and its
eccentric structure, we assume that the center of mass $O$ is at a
distance $a$ away from its geometric center $C$.
We denote coordinates $(X, Y, Z)$ as the space-fixed frame of the
top while ($x$, $y$, $z$) as the body-fixed frame. We adopt $(\phi,
\theta, \psi)$ as the usual Euler angles relative to the space-fixed
frame so that $\Omega=\dot{\phi}$. In addition, we choose the axes
$OY$ and $Oy$ such that they coincide for all time. Therefore, the
space-fixed frame and the body-fixed frame are different only
through the rotation of the nutation angle $\theta$. Furthermore, we
suppose that the spinning about the space-fixed frame is
$\boldsymbol{\Omega}(t)=(0,0,\Omega)$ and the spinning about the
body-fixed frame as $\omega$ where $\Omega$ denotes angular
frequency about the $Z$-axis, and the nano top is assumed to be
initially spinning about the axis $Oz$ with an angular frequency
$\omega$. The distance between the surface and the center of mass is
$h(\theta)=R-a\cos\theta$ and the position vector $OP$ is
$\boldsymbol{r}=(a\sin\theta, 0, -h(\theta))$.  From the above
quantities,  the total angular velocity $\boldsymbol{n}$ and the
total angular momentum $\boldsymbol{L}$ of the nano tippe top's
system in the body-fixed frame are given respectively by
\begin{eqnarray}
&& \boldsymbol{n}=((\omega - \Omega\cos\theta)\sin\theta,
\dot{\theta},
\Omega\sin^{2}\theta+\omega\cos\theta),\nonumber\\
&&
\boldsymbol{L}=((C\omega-A\Omega\cos\theta)\sin\theta,A\dot{\theta},
A\Omega\sin^{2}\theta+C\omega\cos\theta), \nonumber\\
&& \label{1.1}
\end{eqnarray}
where $(A, A, C)$ denote the principal moments of inertia at the
center of mass $O$. In addition, the translational velocity at the
contact point $P$ with respect to $O$ can be written as
$\boldsymbol{v}_{r_{P}}=\boldsymbol{n}\times\boldsymbol{r}$ and
hence the sliding velocity at $P$ is given by
$\boldsymbol{v}_{p}=\boldsymbol{\dot{X}}+\boldsymbol{v}_{r_{P}}$,
where $\boldsymbol{\dot{X}} =(u_{x}, u_{y}, u_{z})$ denotes the
velocity of the center of mass of the nano top. The equations of
motion of the nano top can then be determined for the six degrees of
freedom motion comprising three rotational equations derived from
the Euler equations and three translational equations derived from
Newton second law, namely
\begin{eqnarray}
&& \frac{\partial \boldsymbol{L}}{\partial t} +
\boldsymbol{\Omega}\times\boldsymbol{L}=\boldsymbol{r}\times(\boldsymbol{N}+\boldsymbol{F}),
\nonumber \\ &&
m(\boldsymbol{\ddot{X}}+\boldsymbol{\Omega}\times\boldsymbol{\dot{X}})=\boldsymbol{N}+\boldsymbol{F}+\boldsymbol{W},
\label{1.2}
\end{eqnarray}
where $m$, $\boldsymbol{N}$, $\boldsymbol{F}$, $\boldsymbol{W}$
denote the mass, the normal force, the retarding force and the
weight of the nano top, respectively. Explicit forms of these
equations of motion are given in Appendix A of Ueda \emph{et al.}
\cite{Ueda} and they are also briefly stated in the appendix of this
paper.
%
%
%
%
Next, we  determine a suitable form of the frictional force for the
proposed nano tippe top system. Various theoretical \cite{Dedkov1,
Dedkov2} and molecular dynamics studies \cite{Family, Servantie}
suggest that under the low velocity limit, the frictional force
between two molecules is linearly proportional to their relative
velocity. In addition, Heo \emph{et al.} \cite{Heo} propose that the
frictional force between a fullerene and a carbon nanotube is also
proportional to the fullerene's normal reaction, namely
$|\boldsymbol{F}_{f}|=\mu N$. Further,  molecular dynamics
simulations of Heo \emph{et al.} \cite{Heo} show that the frictional
coefficients for various nanostructures, namely  single-walled
carbon nanotubes, nanopeapods and double walled carbon nanotubes,
under a low pressure regime, are essentially the same, i.e.
$\mu=0.13$. In this paper, we also incorporate a retarding magnetic
force to the model and  we propose that the total retarding force of
the nano top inside the carbon nanotube is given by
\begin{equation}
\boldsymbol{F}=-\boldsymbol{F}_{f}-\boldsymbol{H}(B)=-\frac{0.13N}{\mid\boldsymbol{v}_{P}\mid}\boldsymbol{v}_{P}-\boldsymbol{H}(B),\label{1.3}
\end{equation}
where $\boldsymbol{H}(B)$ denotes the retarding magnetic force
acting at $P$ and $\boldsymbol{v}_{P}$ is the velocity at the
contact point $P$.

\section{Numerical results and discussion}

A fourth order Runge-Kutta method \cite{Burden} is adopted here to
numerically solve this system of six ordinary differential
equations, namely Eq. (\ref{1.2}). Ueda \emph{et al.} \cite{Ueda}
show that in macro scale the top's inversion is strongly dependent
on the ratio of two principal moments of inertia at the center of
mass. Therefore, we check whether the inversion of the  nano top is
possible by running the numerical scheme with various values of
$C/A\in(0,1)$, but we find that this ratio does not effect our
numerical results. However, in all cases a sufficiently large
initial spinning of the nano top is required. Therefore, we choose
the following physical parameters for the numerical iteration:
$R=3.55$ {\AA}, $a=0.1R$, $m=1.196\times10^{-24}$ kg,
$A=(2/3)mR^{2}$, $C=0.5A$  and $H_x=H_y=0.1$ zN  with the following
initial conditions: $\theta=0.1$, $\Omega=0$, $\dot{\theta}=0$,
$\omega=100$ and $\boldsymbol{\dot{X}}=(0,0,0)$. These initial
conditions can be interpreted  that we release the nano top with its
initially spinning (100 Hz) about the $z$-axis, having 0.1 rad
deflection from the $Z$-axis and zero sliding velocity at the
contact point $P$. We use 500 grid points to carry out the numerical
iteration and the numerical results obtained for $\theta$, $\Omega$
and $\omega$ are illustrated in Figs. \ref{fig:Figure2},
\ref{fig:Figure3} and \ref{fig:Figure4}, respectively.
\begin{figure}[!th]
\centering
    \includegraphics[width=11cm,height=7.5cm]{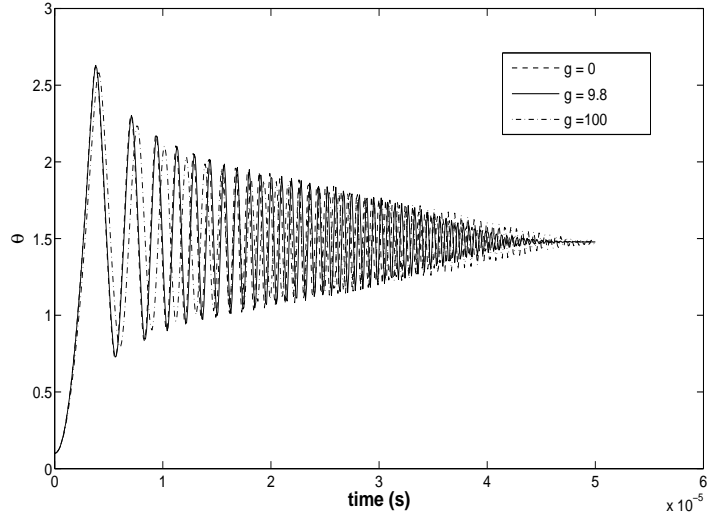}
  \caption{\footnotesize{Nutation angle $\theta$ for $g=0$, 9.8 and 100 ms$^{-2}$ during precession}}
  \label{fig:Figure2}
\end{figure}
\begin{figure}[!th]
\centering
    \includegraphics[width=11cm,height=7.5cm]{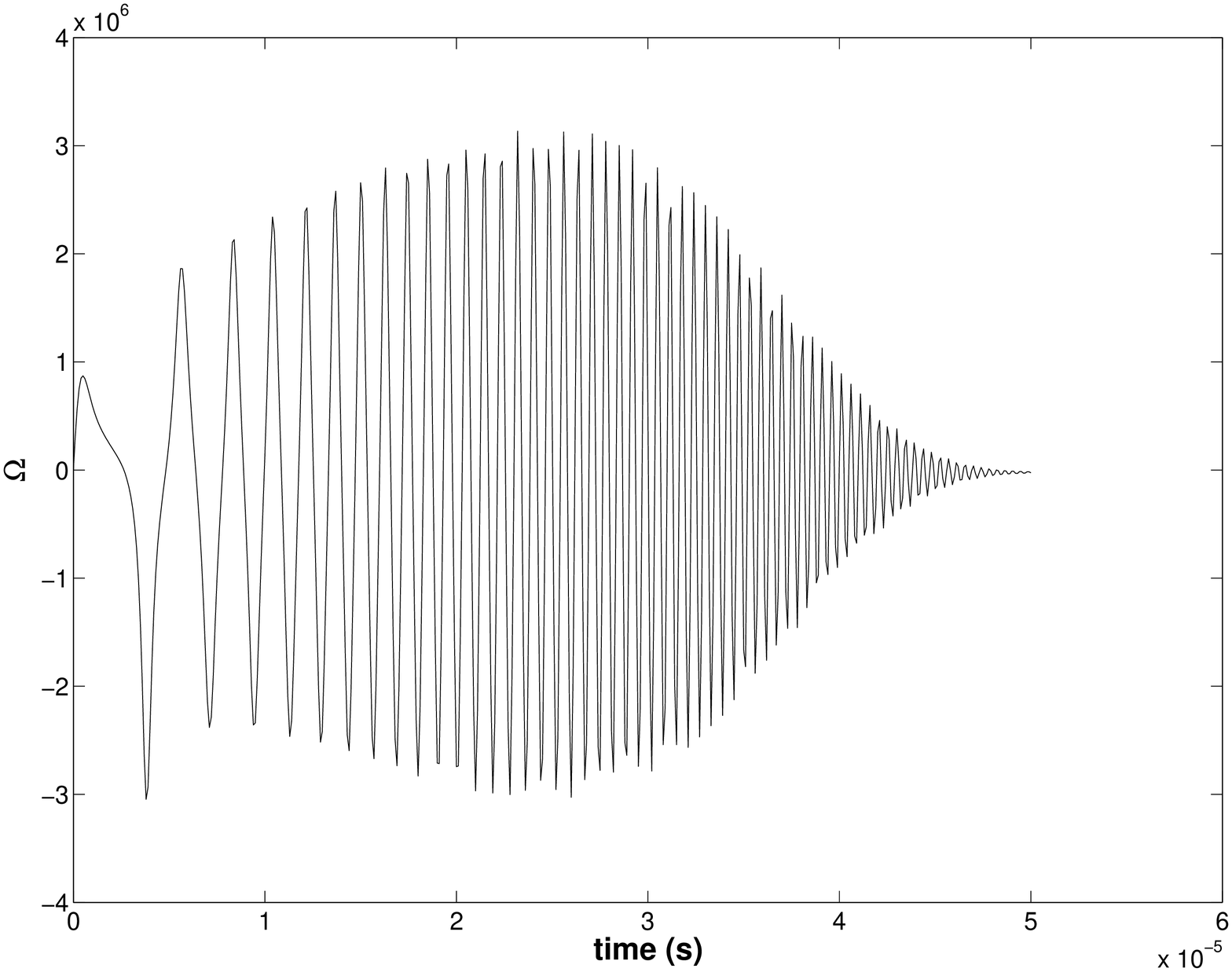}
  \caption{\footnotesize{Angular frequency $\Omega$ about the $Z$-axis during precession}}
  \label{fig:Figure3}
\end{figure}
\begin{figure}[!th]
\centering
    \includegraphics[width=11cm,height=7.5cm]{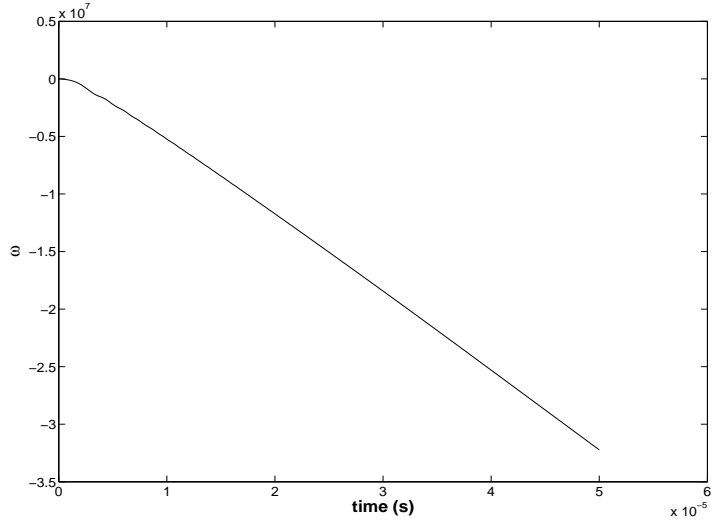}
  \caption{\footnotesize{Angular frequency $\omega$ about the $z$-axis during precession}}
  \label{fig:Figure4}
\end{figure}

Under the total retarding force, which comprises both the frictional
and the magnetic forces at the point of contact $P$,
Fig.~\ref{fig:Figure2} shows that the nano top precesses from its
standing up configuration ($\theta=0$) during the first 5 $\mu$s but
asymptotically approaches the lying down configuration
($\theta=\pi/2$) after $50$ $\mu$s with the decrease in its
oscillating amplitude. During the precession, as shown in
Fig.~\ref{fig:Figure3}, the angular frequency $\Omega$ about the
$Z$-axis increases dramatically in the first micro second indicating
the sudden drop down of the nano top but oscillating around zero
while the nano top has laid down. On the other hand, from
Fig.~\ref{fig:Figure4} the angular frequency $\omega$ about the
$z$-axis also drops from its initial positive value to zero and then
monotonically increases in magnitude to 30 MHz but in the opposite
direction with respect to the direction of the initial spinning.
This implies that the nano top reverses its spinning direction and
gains its spinning speed due to the opposite motion and the energy
gained from the retarding magnetic force respectively. We also note
that the center of mass of the nano top moves in the $X,
Y$-directions only resulting in sliding friction but remains intact
in the $Z$-direction (see Figs.~\ref{fig:Figure10} and
~\ref{fig:Figure11}). The whole precession process is illustrated in
Fig.~\ref{fig:Figure5}.
\begin{figure}[!]
\centering
    \includegraphics[width=8cm,height=2.2cm]{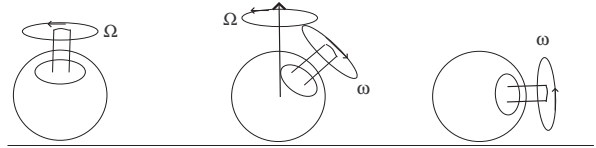}
  \caption{\footnotesize{Precession of nano tippe top}}
  \label{fig:Figure5}
\end{figure}
Unlike the classical tippe top, we observe no flip over phenomenon
in the nano top's precession. This no inversion phenomenon arises
from the fact that the gravitational force is negligibly small  at
the nano scale. As confirmed by Fig.~\ref{fig:Figure2}, the
precession is not effected by the gravitational force since the
numerical results for  $g=0$ and 9.8 ms$^{-2}$ almost coincide with
each other, and the same behaviour is obtained for $g=100$
ms$^{-2}$.

Once the nano top is spinning about the lying down axis, it is
important to determine a possible way for which the nano top can
retract to its initial standing up position. We find that the nano
top will retract smoothly back to its standing up axis upon applying
a magnetic force of the same magnitude but in the opposite direction
to the previous retarding magnetic force and the numerical solution
for $\theta$ is shown in Fig.~\ref{fig:Figure6}.
\begin{figure}[!th]
\centering
    \includegraphics[width=11cm,height=7.5cm]{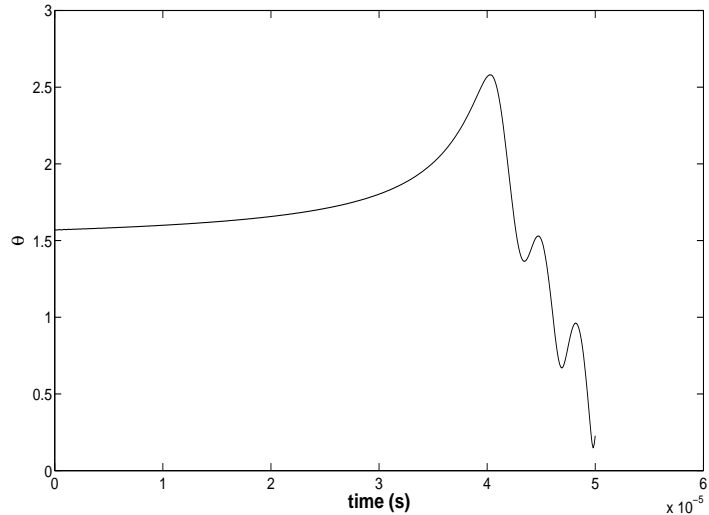}
  \caption{\footnotesize{Nutation angle $\theta$} after applying the reversed retarding magnetic field at $P$}
  \label{fig:Figure6}
\end{figure}

Accordingly, by adjusting the magnetic field, we can manually
control the nano top's switching between the standing up  and the
lying down states and hence it can be utilized as a nano-computing
memory device. Moreover, the main advantages of utilizing the nano
tops as a memory device are that it possesses a remarkably short
relaxation time ($\approx$ 50 $\mu$s) resulting in a higher
computational speed, and it is small in size and hence provides a
larger memory capacitance. Most importantly, it is simpler to
control as compared with the electron-spin quantum memories.
Finally, the integration of the self-assembled hybrid nanostructure
known as nanopeapods  and the ideas of nano tippe top developed here
may lead to practical computing memory devices, which currently
require large numbers of bit handling.

\section{Conclusion}

A nano tippe top formed from a C$_{60}$ fullerene and spinning
either on a graphene sheet or inside a carbon nanotube is
investigated as a possible candidate for a computing memory device.
The equations of motion for such a nano top are described and we
find that while the retarding magnetic force makes the nano top
precess, it does not flip over as in the classical tippe top, but
due to the fact that gravity is negligible at the nano scale, it
adopts a lying down position. In addition, while the nano top is in
the lying down position, if we apply the magnetic force which is of
the same magnitude but in the opposite direction to the previous
retarding magnetic force, then the nano top will return to its
standing up position. Hence, the standing up and lying down
configurations of the nano top might be considered as two bit
states, which gives rise to their potential use as a future memory
device.

\pagestyle{myheadings}
\section*{Acknowledgements}
The authors are grateful for the Australian Research Council for
support through the Discovery Project Scheme and the provision of an
Australian Postdoctoral Fellowship for NT and an Australian
Professorial Fellowship for JMH.

\appendix
\section{Equations of motion}

Here, we briefly state explicit forms of equations of motion
obtained from Eq.~(\ref{1.2}). From Eq.~(\ref{1.2})$_{1}$, the three
rotational equations are given by
\begin{eqnarray}
&& A\ddot{\theta}=-\Omega(C\omega-A\Omega\cos\theta)\sin\theta -
aN\sin\theta - h(\theta)F_{x},\nonumber\\
&&
A\sin\theta\dot{\Omega}=(C\omega-2A\Omega\cos\theta)\dot{\theta}+(a-R\cos\theta)F_{y},\nonumber\\
&& C\dot{\omega}=R\sin\theta F_{y}, \label{A1}
\end{eqnarray}
where $h(\theta) = R - a\cos\theta$.  From Eq.~(\ref{1.2})$_{2}$ the
three translational equations are of the form
\begin{eqnarray}
&& m\dot{u}_{x}=m\Omega u_{y} + F_{x}, \nonumber\\
&& m\dot{u}_{y}=-m\Omega u_{x} + F_{y}, \nonumber\\
&& m\dot{u}_{z}=N-mg,\label{A2}
\end{eqnarray}
where
\begin{eqnarray}
 F_{x}&=&\frac{-\mu N \left\{ u_{x}-h(\theta)\dot{\theta}
\right\}}{\sqrt{ \left\{ u_{x}-h(\theta)\dot{\theta} \right\}^{2} +
\left\{ u_{y} + \left[
R(\omega-\Omega\cos\theta)+a\Omega\right]\sin\theta\right\}^{2}
}}\nonumber\\
&& - H_{x}, \nonumber \\
 F_{y}&=&\frac{-\mu N \left\{ u_{y}+ \left[ R(\omega -
\Omega\cos\theta) +a\Omega \right]\sin\theta \right\}}{\sqrt{
\left\{ u_{x}-h(\theta)\dot{\theta} \right\}^{2} + \left\{ u_{y} +
\left[
R(\omega-\Omega\cos\theta)+a\Omega\right]\sin\theta\right\}^{2}
}}\nonumber\\
&& - H_{y},\label{A3}
\end{eqnarray}
and $H_{x}$ and $H_{y}$ denote the strength of the retarding
magnetic force in the $x$- and $y$-directions respectively. By
multiplying $\sin\theta$ both sides of Eq.~(\ref{A1})$_{2}$ we
obtain
\begin{equation}
A\sin^2\theta \dot{\Omega} = (C\omega -
2A\Omega\cos\theta)\sin\theta\dot{\theta} +
(a-R\cos\theta)F_y\sin\theta. \label{A4}
\end{equation}
From  Eq.~(\ref{A1})$_{3}$ we have $F_y\sin\theta =
C\dot{\omega}/R$, which upon substituting into Eq.~(\ref{A4}) we
have
$$
A\sin^2\theta \dot{\Omega} + 2A\Omega\cos\theta\sin\theta
\dot{\theta} = C\omega\sin\theta\dot{\theta} - C\dot{\omega}
\cos\theta+ aC\dot{\omega}/R,
$$
which can be written as
\begin{equation}
A\frac{d(\Omega\sin^2\theta)}{dt} =  \frac{aC}{R}\frac{d\omega}{dt}
- C\frac{d(\omega\cos\theta)}{dt}. \label{A5}
\end{equation}
Thus, from Eq.~(\ref{A5}) we obtain
\begin{equation}
A\Omega \sin^2\theta= aC\omega/R - C\omega\cos\theta + J^*,
\label{A6}
\end{equation}
where $J^*$ is a constant of integration. By multiplying $R$ both
sides of Eq.~(\ref{A6}) we have
\begin{equation}
J = A\Omega R\sin^2\theta + C\omega(R\cos\theta -a), \label{Jellet}
\end{equation}
where the constant $J$ is referred to as the Jellett's constant
\cite{Ueda} and our numerical result indicates that this constant is
zero throughout the precession process.

 As a  benchmark, we use our numerical scheme to show the behaviour
of the classical macro scale tippe top with zero magnetic force.
Here, the values of parameters are taken to be $R=0.015$ m, $a=0.1R$
m, $M=0.015$ kg, $A=C=(2/5)MR^{2}$ and $\mu=0.1$, with initial
conditions: $\theta=0.1$, $\omega=100$, $\Omega=0$ and
$\boldsymbol{\dot{X}}=(0,0,0)$. Thus, from Eqs.~(\ref{A2}) and
(\ref{A3}) we obtain numerical results for nutation angle
$\theta(t)$ as illustrated graphically in Fig.~\ref{fig:Figure7}.
From this figure, we can see that $\theta$ approaches $\pi$ implying
that the top flips over, which is consistent with Ueda~{\em et al.}
\cite{Ueda}.
\begin{figure}[!th]
\centering
    \includegraphics[width=11cm,height=7.5cm]{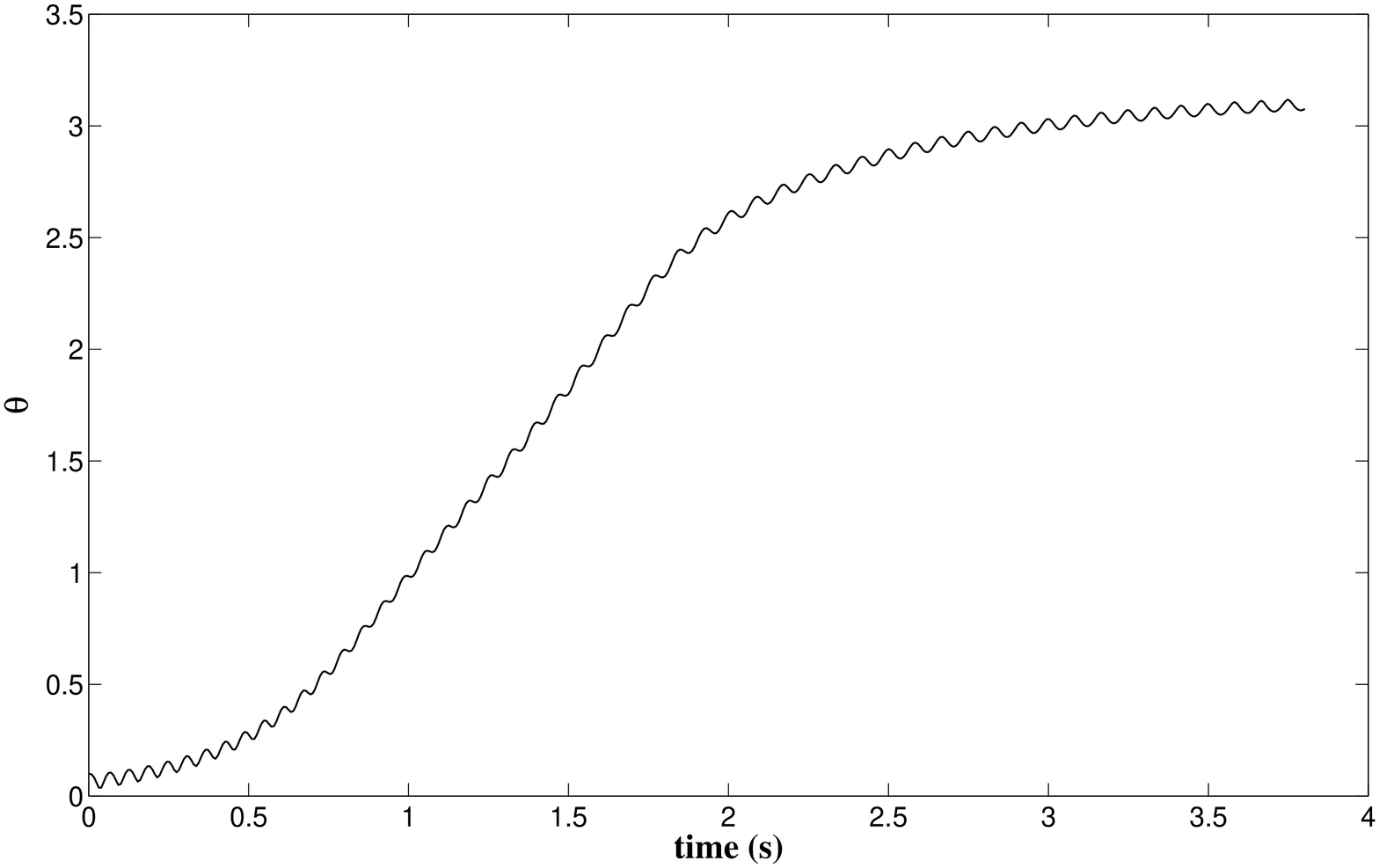}
  \caption{\footnotesize{Nutation angle $\theta$ for classical macro scale tippe top}}
  \label{fig:Figure7}
\end{figure}

\section{Stability analysis}

In this section, we use a simple stability argument to investigate
the stability of the nano top for  $\theta =0$, $\pi/2$ and $\pi$.
Without the external magnetic retarding force, i.e. $H_{x}=H_{y}=0$,
we consider a small perturbation around the standing up axis, namely
$\theta=\epsilon$ for small $\epsilon>0$. Eq.~(\ref{A1})$_{1}$
becomes
\begin{equation}
\ddot{\epsilon} = -aN\epsilon/A, \label{B1}
\end{equation}
which implies that the spinning nano top is always stable. If we
incorporate the external magnetic retarding force, i.e.
$H_{x}=H_{y}=H$, which is a constant. For $\theta=\epsilon$,
Eq.~(\ref{A1})$_{1}$ becomes
\begin{equation}
\ddot{\epsilon} = -aN\epsilon/A+(R-a)H/A, \label{B2}
\end{equation}
which also implies that the nano top is stable. We note that for the
case $\theta = \epsilon$, we assume $\Omega =0$.

Next, we  check the stability of the nano top at $\theta=\pi$. Upon
substituting $\theta=\pi-\epsilon$, Eq.~(\ref{A1})$_{1}$ becomes
\begin{equation}
\ddot{\epsilon}=\frac{(C+A)\Omega^2 + aN}{A}\epsilon -
\frac{(R+a)H}{A}, \label{B3}
\end{equation}
noting that  $\omega = \Omega$ when $\theta $ tends to $\pi$.
Eq.~(\ref{B3}) implies that the nano top is always unstable, and
therefore it does not flip over.

For the lying down position, i.e. $\theta=\pi/2-\epsilon$,
Eq.~(\ref{A1})$_{1}$ becomes
\begin{equation}
\ddot{\epsilon}=-\frac{A\Omega^2-aH}{A}\epsilon - \frac{HR -
C\omega\Omega - aN}{A},\label{B4}
\end{equation}
which is stable if $A\Omega^2 - aH >0$ or $\Omega^2 > aH/A$.

\section{Asymptotic expansion for~$\theta=\pi/2$ }

In this section, we check the compatibility between our numerical
results and the asymptotic expansions for the governing ordinary
differential equations given in Eq.~(\ref{1.2}) or Eq.~(\ref{A1})
when the nano top is lying down and the time $t$ is sufficiently
large. Upon substituting $\theta=\pi/2$ into Eqs.~(\ref{A1})$_{1}$,
~(\ref{A1})$_{3}$, ~(\ref{A2})$_{1}$ and ~(\ref{A2})$_{2}$, we
obtain
\begin{eqnarray}
 && \Omega= RH/(C\omega),  \quad
 \dot{\omega}=-RH/C, \nonumber\\
 && m\dot{u_{x}}=m\Omega u_{y}-H,  \quad  m\dot{u_{y}}=-m\Omega
u_{x}-H, \label{C1}
\end{eqnarray}
respectively. Noting here that we assume $H_x = H_y =H$. From
Eq.~(\ref{C1})$_{2}$, upon integrating both sides by $t$, we have
\begin{equation}
\omega=-RHt/C+c_{1}. \label{C2}
\end{equation}
This asymptotic expansion and its corresponding numerical result for
$\omega$ are plotted together in Fig.~\ref{fig:Figure8}. We note
that the constants used here are given by $R=3.55$ {\AA}, $C=0.5A$,
$A=(2/3)mR^{2}$, $m=1.196\times10^{-24}$ kg  and $H=0.1$ zN. We fit
Eq.~(\ref{C2}) with the numerical solution for $\omega$ and obtain
$c_{1}=3\times10^{6}$.
\begin{figure}[!h]
\centering
    \includegraphics[width=11cm,height=7.5cm]{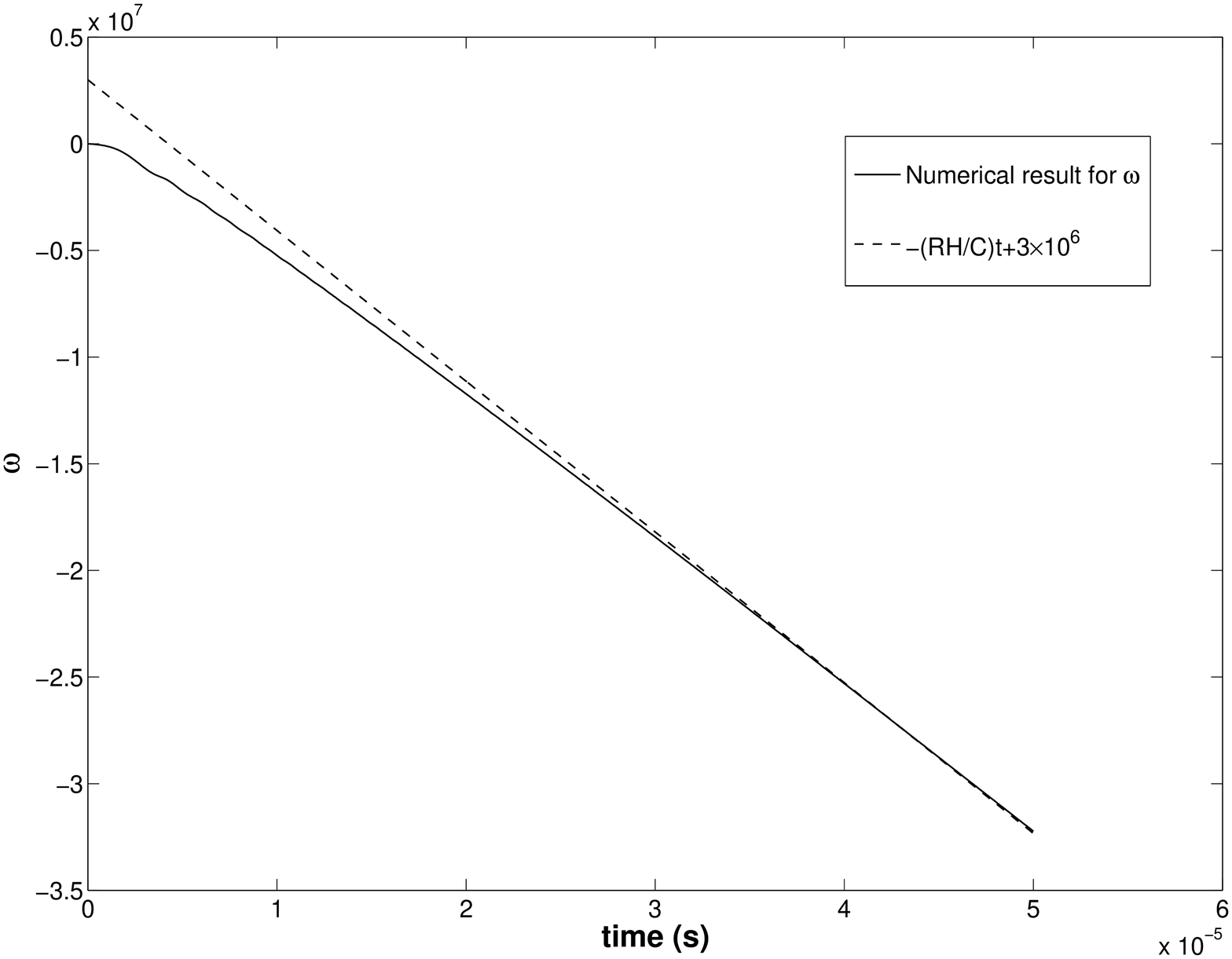}
  \caption{\footnotesize{Asymptotic expansion for $\omega$}}
  \label{fig:Figure8}
\end{figure}
Upon knowing $\omega$, we can  determine $\Omega$ by utilizing
Eq.~(\ref{C1})$_{1}$ as
\begin{equation}
\Omega=\frac{RH}{c_1C-RHt},\label{C3}
\end{equation}
where this asymptotic equation decays to zero for a sufficiently
large $t$ and the corresponding result with its numerical solution
are plotted in Fig.~\ref{fig:Figure9}.
\begin{figure}[!h]
\centering
    \includegraphics[width=11cm,height=7.5cm]{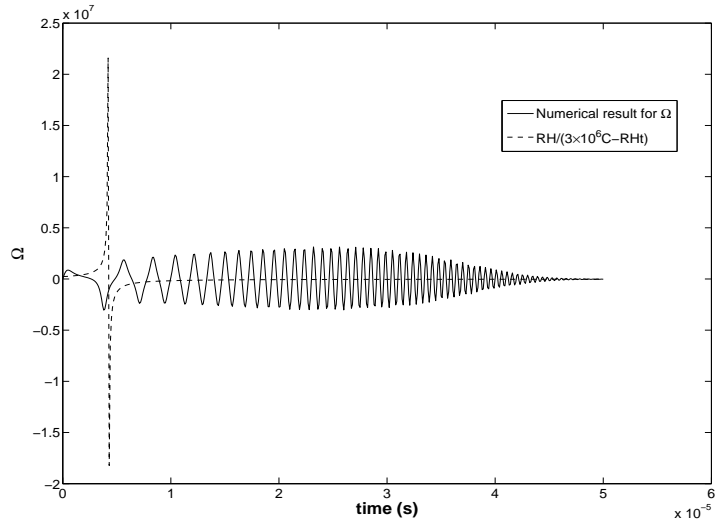}
  \caption{\footnotesize{Asymptotic expansion for $\Omega$}}
  \label{fig:Figure9}
\end{figure}
Given the asymptotic expansions for $\Omega$, we can determine the
asymptotic expansions for both  $u_{x}$ and $u_{y}$, which are given
by
\begin{eqnarray}
&& m\dot{u_{x}}=\frac{mRH}{(c_{1}C-RHt)}u_{y}-H, \nonumber\\
&& m\dot{u_{y}}=-\frac{mRH}{(c_{1}C-RHt)}u_{x}-H. \label{C4}
\end{eqnarray}

We can solve Eq.~(\ref{C4}) analytically by introducing $v=u_x +
iu_y$, where $i=\sqrt{-1}$. By multiplying  $i$ both sides of
Eq.~(\ref{C4})$_{2}$ and combining  with Eq.~(\ref{C4})$_{1}$ we
have
\begin{equation}
\dot{v} + \frac{RHi}{\left(c_1C - RHt \right)}v = -\frac{H}{m}(1+i).
\label{v_dot}
\end{equation}
By multiplying both sides of Eq.~(\ref{v_dot}) by an integrating
factor $\exp\left(-i\log\left(c_1C-RHt\right)\right) =
1/(c_1C-RHt)^i$ we obtain
$$\frac{d}{dt}\left( \frac{v}{\left(c_1C-RHt\right)^i} \right) = -\frac{H}{m}\frac{(1+i)}{(c_1C - RHt)^i},$$
which on integrating we deduce
$$
\frac{v}{\left(c_1C - RHt\right)^i}= \frac{(1+i)}{(1-i)}\frac{1}{Rm}
\left(c_1C - RHt\right)^{1-i} + c_2,
$$
where $c_2$ denotes a constant. Thus, the solution of
Eq.~(\ref{v_dot}) is given in the form
\begin{equation}
v(t) = \frac{(1+i)}{(1-i)}\frac{(c_1C-RHt)}{Rm} +
c_2e^{i\log(c_1C-RHt)}. \label{v_sol}
\end{equation}
To determine the complex constant $c_2$ to fit our numerical data,
we assign at $t=t^*$, $u_x(t^*) = u_1$ and $u_y(t^*) = u_2$, where
$u_1$ and $u_2$ are constants. Thus, $c_2$ is given by
$$ c_2 = e^{-i\log\lambda}\left\{u_1 + iu_2 - \frac{i\lambda}{Rm} \right\},$$
where $\lambda = c_1C - RHt^*$. Upon  substituting into
Eq.~(\ref{v_sol}) and simplifying we have
$$v(t) = \frac{i(c_1C - RHt)}{Rm} + e^{i\beta(t)}\left\{u_1 +iu_2 - \frac{i\lambda}{Rm} \right\}, $$
where $\beta(t) = \log \left( [c_1C - RHt]/\lambda \right)$. Next,
we expand the above equation  using Euler's formula, $e^{i\beta} =
\cos\beta + i\sin\beta$, to obtain
\begin{eqnarray}
v(t) &=& u_1 \cos\beta(t)+ \left\{ \frac{\lambda}{Rm} -u_2 \right\}
\sin\beta(t) + \frac{i(c_1C-RHt)}{Rm}  \nonumber \\
&& + iu_1\sin\beta(t) - i\left\{\frac{\lambda}{Rm} - u_2 \right\}
\cos\beta(t). \label{v_sol_c2}
\end{eqnarray}
 Since $v = u_x + iu_y$,
Eq.~(\ref{v_sol_c2}) gives rise to  analytical solutions for $u_x$
and $u_y$, namely
\begin{eqnarray}
u_x &=& u_1 \cos\beta(t)+ \left\{ \frac{\lambda}{Rm} -u_2 \right\}
\sin\beta(t), \nonumber
\\
u_y &=& \frac{(c_1C-RHt)}{Rm}  + u_1\sin\beta(t) -
\left\{\frac{\lambda}{Rm} - u_2 \right\} \cos\beta(t). \nonumber \\
\label{ux_uy}
\end{eqnarray}
The solutions (\ref{ux_uy}) for $u_x$ and $u_y$ are plotted together
with their corresponding numerical results in
Figs.~\ref{fig:Figure10} and \ref{fig:Figure11}, respectively. We
note that the values of constants used here are taken to be $t^*
=5\times10^{-5}$ s, $u_1=-6\times 10^{-7}$ ms$^{-1}$ and
$u_2=-4\times 10^{-3}$ ms$^{-1}$. As can be seen from these figures,
the solutions (\ref{ux_uy}) agree with the numerical results for
sufficiently large $t$.
\begin{figure}[!h]
\centering
  \includegraphics[width=11cm,height=7.5cm]{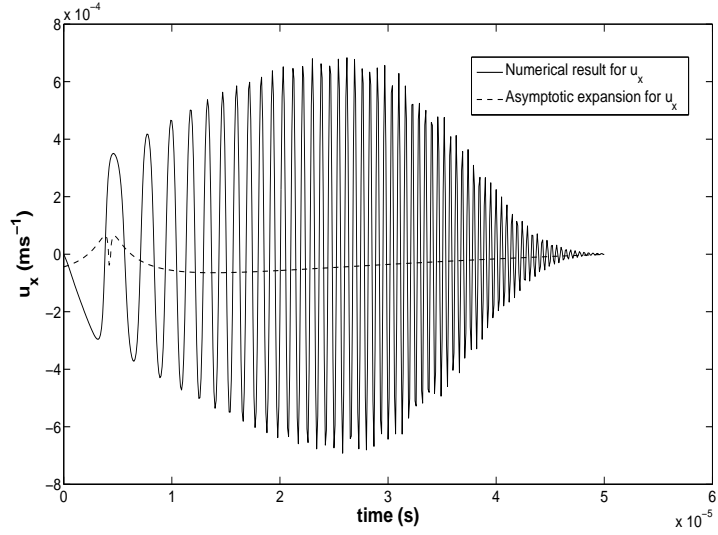}
  \caption{\footnotesize{Asymptotic expansion for $u_{x}$ (\ref{ux_uy})$_1$ in comparison with numerical result}}
  \label{fig:Figure10}
\end{figure}
\begin{figure}[!th]
\centering
    \includegraphics[width=11cm,height=7.5cm]{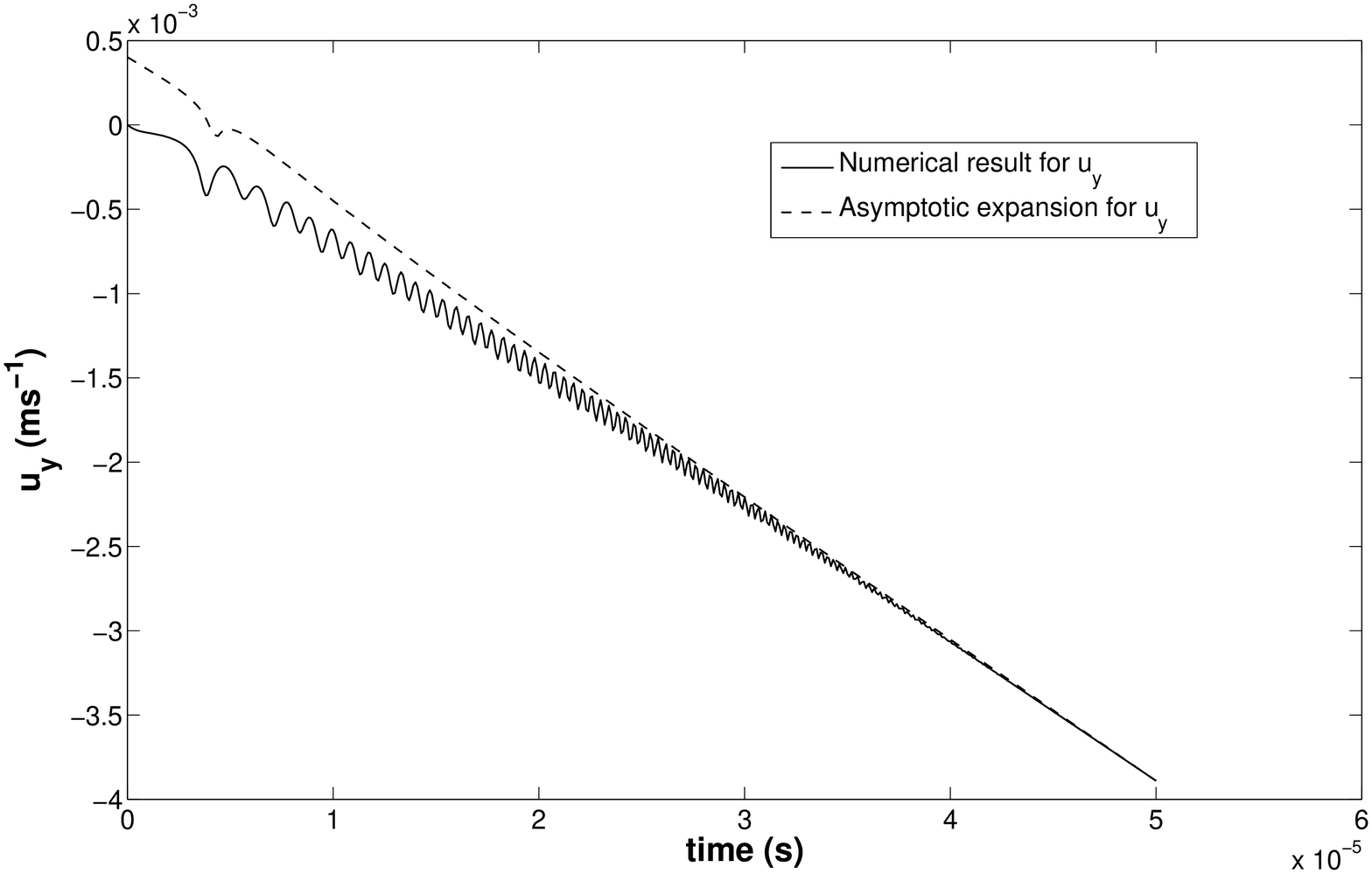}
  \caption{\footnotesize{Asymptotic expansion for $u_{y}$ (\ref{ux_uy})$_2$ in comparison with numerical result}}
  \label{fig:Figure11}
\end{figure}

\section{Retarding magnetic force as step function}

In this section, we investigate the precession of the nano top
subject to a  magnetic force which is applied only in the
$y$-direction and only for a finite time $t_0$. In particular, we
consider $H_y = H\mbox{H}(t_0-t)$, where $H$ is a constant
representing the strength of the magnetic force, $\mbox{H}(t)$ is
the Heaviside unit step function and $t_0$ denotes the time when the
magnetic force is switched off.  Two cases are examined, namely $t_0
= 0.8\times10^{-5}$ and $10^{-6}$ seconds. We observe that in the
former case the nano top precesses from its initial standing up
position to $\theta=2.1$ soon after the retarding magnetic field is
switched off and then it oscillates about $\theta=\pi/2$, as
demonstrated in Fig.~\ref{fig:Figure12}.
\begin{figure}[htbp]
\centering
    \includegraphics[width=11cm,height=7.5cm]{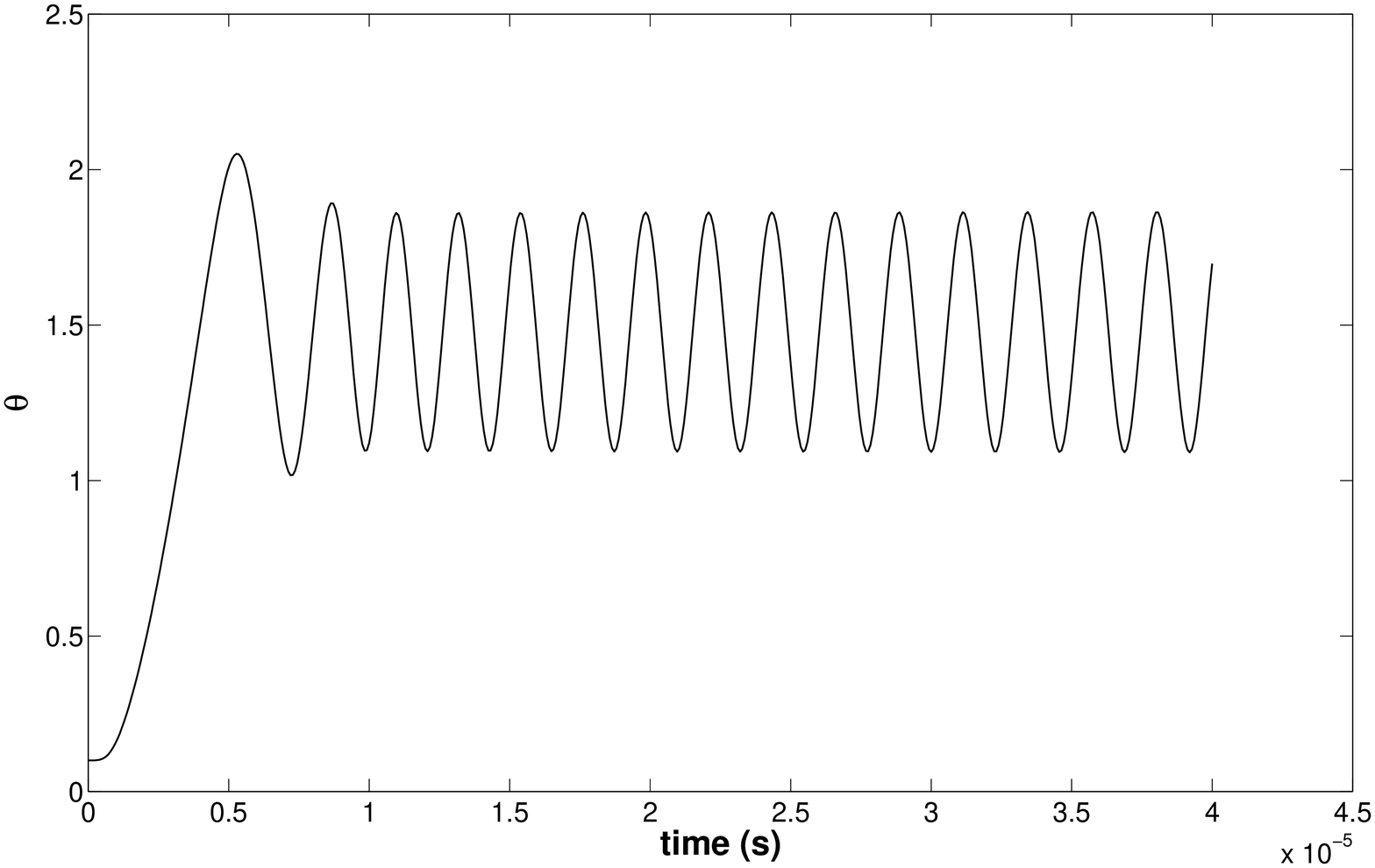}
  \caption{\footnotesize{Nutation angle $\theta$ for nano top when $H_x =0$ and $H_y=H$H($t_0-t$) where $t_0 = 0.8\times10^{-5}$ s}}
  \label{fig:Figure12}
\end{figure}
The effect of the Heaviside function can be seen from the behaviour
of $\omega$, which is shown in Fig.~\ref{fig:Figure13}. From this
figure,  before the switch off time $t_0=0.8\times10^{-5}$, the
magnitude of $\omega$ increases but in the opposite direction with
respect to the direction of the initial spin due mainly to the
retarding magnetic force. After $t=0.8\times10^{-5}$, since the
retarding force is switched off, the magnitude of $\omega$ starts to
decrease by the effect of the frictional force only. For the latter
case, we find from Fig.~\ref{fig:Figure14} that the application of
$H_{y}$ for $10^{-6}$ s does not provide sufficient angular momentum
for the nano top to lie down in a stable configuration. The
variation of the corresponding $\omega$ for this case is presented
in Fig.~\ref{fig:Figure15}.
\begin{figure}[!h]
\centering
    \includegraphics[width=11cm,height=7.5cm]{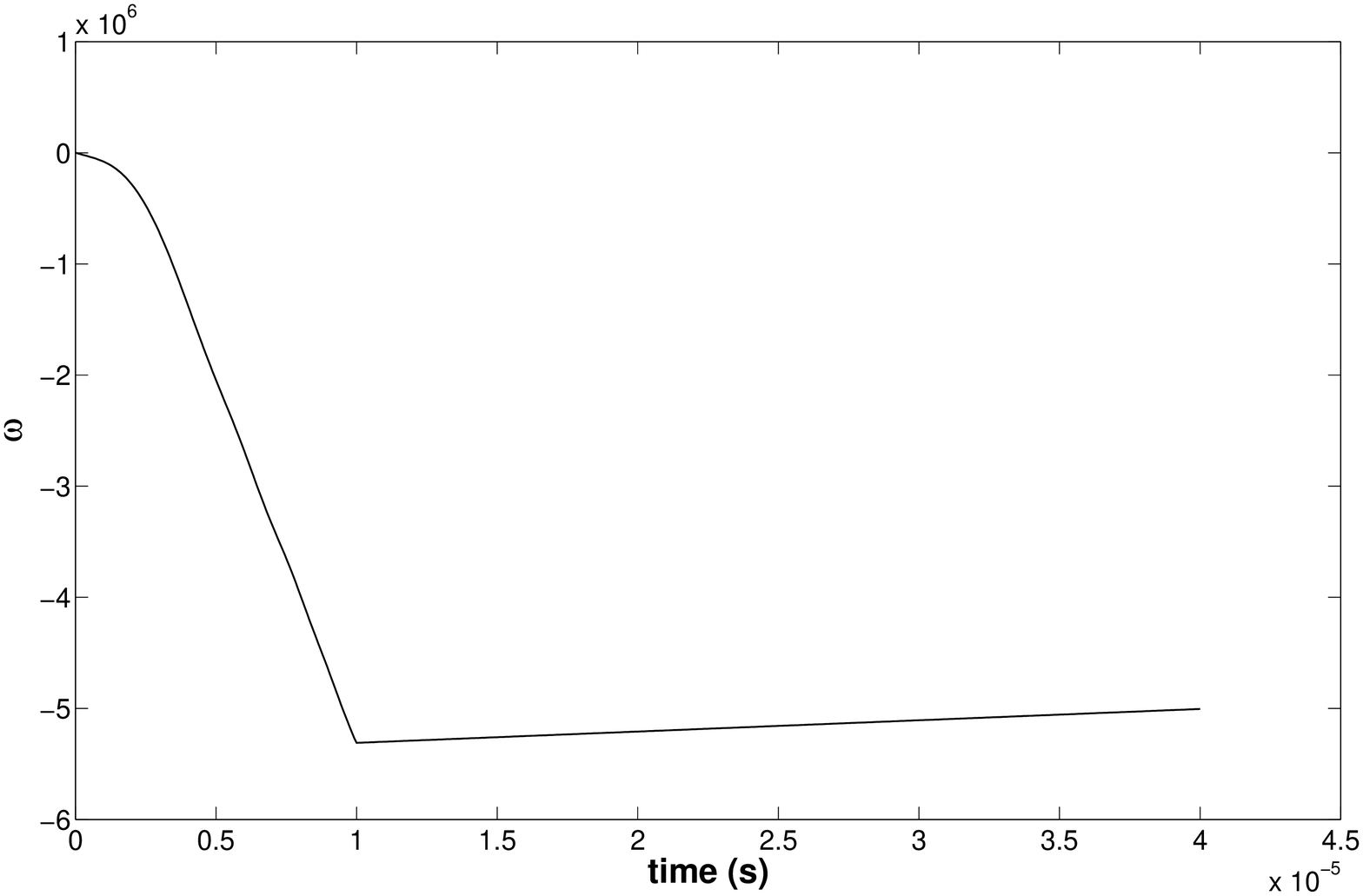}
  \caption{\footnotesize{Angular frequency $\omega$ for nano top when $H_x=0$ and  $H_y=H$H($t_0-t$) where $t_0 = 0.8\times10^{-5}$ s}}
  \label{fig:Figure13}
\end{figure}
\begin{figure}[!h]
\centering
    \includegraphics[width=11cm,height=7.5cm]{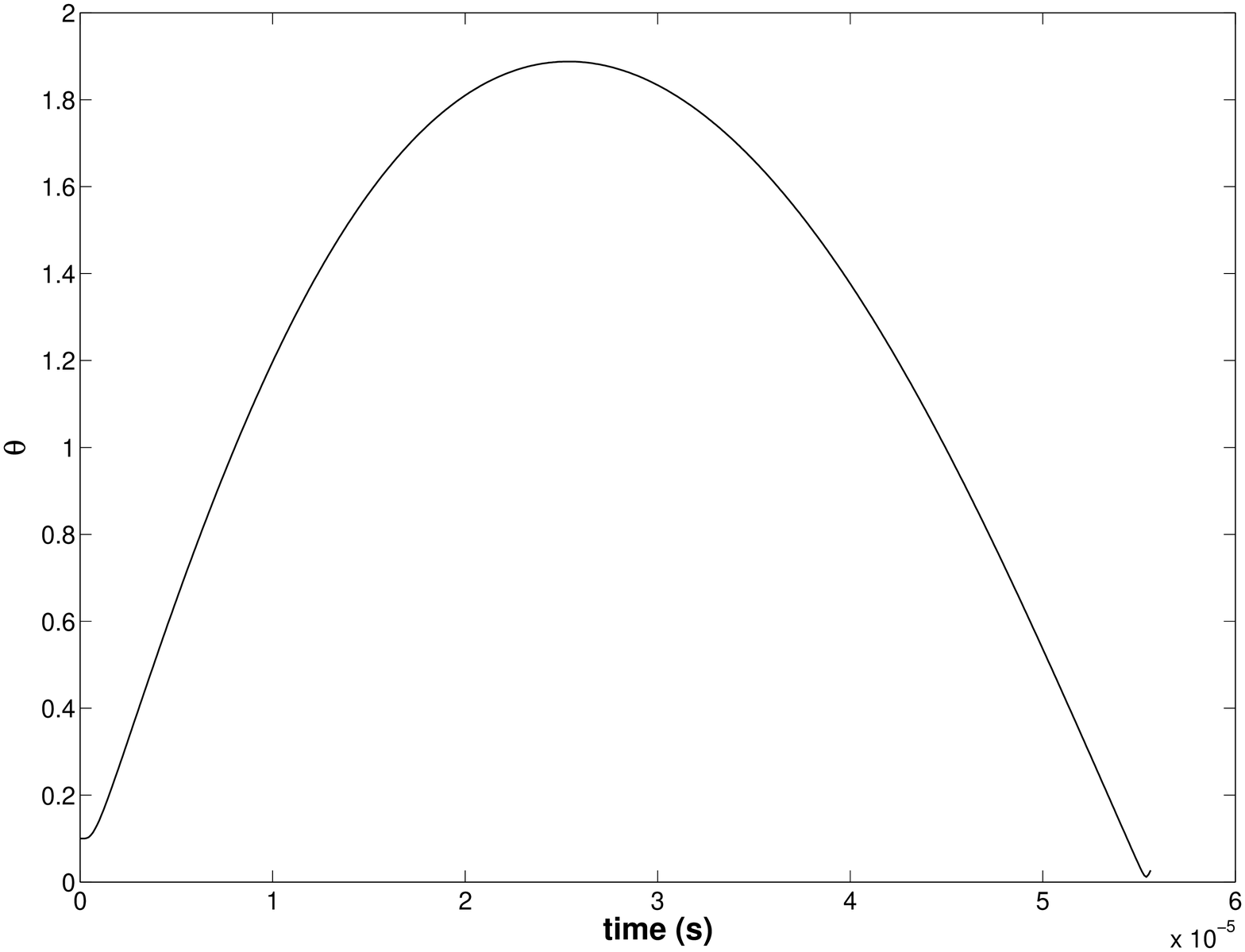}
  \caption{\footnotesize{Nutation angle $\theta$ for nano top when $H_x=0$ and $H_y=H$H($t_0-t$) where $t_0 = 10^{-6}$ s}}
  \label{fig:Figure14}
\end{figure}
\begin{figure}[!h]
\centering
    \includegraphics[width=11cm,height=7.5cm]{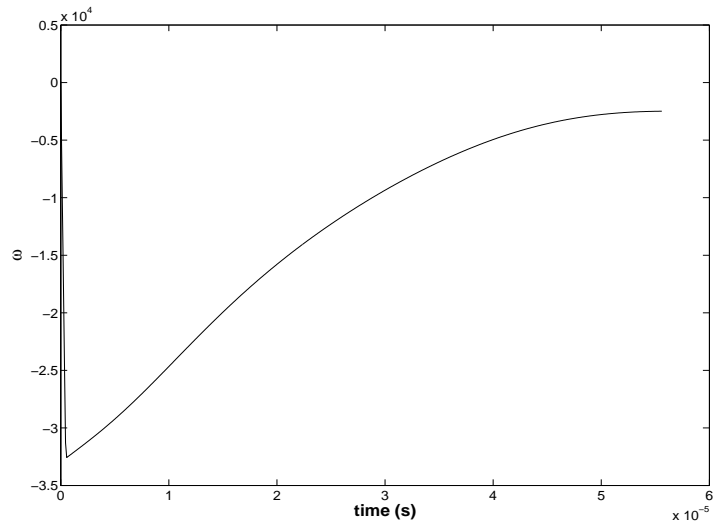}
  \caption{\footnotesize{Angular frequency $\omega$ for nano top when $H_x=0$ and $H_y=H$H($t_0-t$) where $t_0=10^{-6}$ s}}
  \label{fig:Figure15}
\end{figure}

\end{document}